\documentstyle[aps,prl,epsfig,twocolumn]{revtex}
\begin{document}
\twocolumn[\hsize\textwidth\columnwidth\hsize\csname
  @twocolumnfalse\endcsname
\preprint{IMSc-96/12/31}
%\preprint{hep-th@ftp/9612306}
\title{Implications of chiral symmetry for 
scalar isoscalar channel and multi- nucleon forces }
\author{Ramesh Anishetty, N.D. Hari Dass and H.S. Sharatchandra \cite{e-mail}} 
\address{The Institute of Mathematical Sciences, Chennai - 600 113, INDIA}
\maketitle
\begin{abstract}
Chiral perturbation theory has to be substantially modified for the problem 
of nuclear forces.  Loop effects dominate over tree level effects. The dominant 
behaviours in the scalar isoscalar channel and multi-nucleon forces are obtained 
in the chiral limit and for a small explicit breaking.
\end{abstract}
\vskip 2pc] % end \twocolumn[...]
%\pacs{12.39.Fe,13.75.Cs,21.30.+y,13.75.Gx}
In this Letter we point out that chiral symmetry has some clear
model independent implications for the scalar isoscalar 
( referred to as the $\sigma$) channel of the nucleon nucleon interactions and
for multinucleon foces.We obtain dominant effects of all pion exchanges 
between nucleons.
It has been emphasised by Weinberg \cite{wei} that the implications of chiral
symmetry to the nucleon potential is in a different class from the usual
low energy theorems involving pions with each other or scattering off 
a heavy nucleon because there is no such thing as a soft nucleon.
When the effective Lagrangian is used it appears that there is no interesting 
prediction for an expansion in powers of momentum transfer. With the chiral 
transformation of the 
nucleon field defined as in the non-linear realisation \cite{wei-nl}, 
(see  below) $\overline N'N'$ is a chiral invariant.
Therefore two and  multi- nucleon interactions of arbitrary strength are 
allowed by symmetry.This is in contrast to the scattering of pions with each 
other or with other chiral matter where an expansion in powers of momenta 
is sought in low energy theorems \cite{wei-nn} and chiral perturbation 
theory \cite{cpt}.

We argue that the implication of spontaneously broken chiral symmetry
for the nuclear forces is elsewhere.  Consider the spin zero isospin one 
channel of the nucleon nucleon scattering in the chiral limit.
There is a charecteristic dipole-dipole type of long range interaction 
as a consequence of the massless pion exchanged.With an explicit
breaking of chiral symmetry, the dominant effect is to change the long  range 
into the Yukawa interaction.In the same way 
it is more appropriate to isolate the dominant (qualitative and 
quantitative) effect in the $\sigma$ channel in the chiral limit and
for a small explicit breaking.  We do this here.The implications of 
chiral symmetry for the nuclear forces has attracted considerable attention 
recently \cite{cel}-\cite{kap}.In contrast, we point out that chiral peturbation 
theory has to be modified substantially for the NN interactions in the 
$\sigma$ channel as 
the loop effects dominate over the tree level effects.We also emphasise the
crucial role played by exlicit chiral symmetry breaking and the pion nucleon $\sigma$ 
term.

Our starting point is 
that not only the transverse but also the longitudinal susceptibility  
diverges  in a ferromagnet.The implication is that the $\sigma$
propagator  has an unusual though universal logarithmic infrared singularity 
\cite{rg} \cite{abhs}.To see this in the linear $\sigma$ model, we need
an infinite summation of graphs \cite{ahs}, an example of which is the 
$N \rightarrow \infty$ approximation \cite{abhs}.
(This only refers to the infrared behaviour
and does not rule out a pole at some $q^2 \neq 0$.) 

$\sigma$ does not appear as a fundamental field in the non-linear
$\sigma$-model but it is the chiral partner of the pions and therefore has a
physical meaning as a composite field.
Its correlation  refers to the longitudinal susceptibility in a ferromagnet
which is as important as the transverse susceptibilty.
It is easier to understand the logarithmic infrared singularity here.
In a O(N) non-linear $\sigma$-model we have
$\sigma^2 + \vec \pi^2= f^2$, so that
$\sigma' = -:\vec \pi^2:/(2f) + :\vec \pi^4:/(8f^3)+\cdots$
where $\sigma' = \sigma -<0|\sigma|0>$.
Using the massless propagator for $\vec \pi$,  we get (Fig.1a), 
\begin{equation}
<\sigma'(-{\bf q}) \; \sigma'({\bf q})>  \sim  ln \,(\Lambda^2 / \bf q^2),
\;q \rightarrow 0  
\label{sigir}
\end{equation}
where $\Lambda$ is the microscopic scale (or a cutoff ) in the theory.
We will refer to this as the {\it logarithmic enhancement in the $\sigma$ 
channel}.  Application \cite{rg} of  the renormalization group shows 
that the {\it leading} infrared behaviour Eqn. \ref{sigir} is exact.
The pion loop (Fig. 1a)  plays a special role because it is the only infrared 
singular graph.  Thus $\sigma'$ behaves like a composite $ :\vec \pi^2: $ of 
two pions {\it in the infrared}\cite{rg}.

\begin{figure}[htb]
\begin{center}
\mbox{\epsfig{file=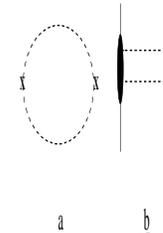,width=3truecm,height=2truecm,angle=-90}}
\caption{ a.Log enhancement in the $\sigma$ channel.
b.NN interaction due to a two pion exchange.}
\label{Fig 1.}
\end{center}
\end{figure}
Chiral perturbation theory would indirectly include the effects of the 
infrared behaviour of the $\sigma'$ as an one loop effect.Indeed
the ln effect Eqn. \ref{sigir}  appeared first \cite{leh} when the 
soft pion amplitudes were corrected for unitarity.
We emphasise that it survives as the dominant effect to all orders.
Note that chiral perturbation theory orders the graphs according to the
powers of momenta and does not count the  logarithms.  As the logarithm is 
all important now, this phenomenon should be kept track of explicitly.

We now consider the dominant effect of all pion exchanges between 
nucleons. It  is possible to handle this 
at low momentum transfers because the amplitude for emission of any 
number of soft (virtual)  pions from a nucleon is given by the
soft pion theorems.Consider two pion exchanges (Fig.1b) to
be specific. In the {\it chiral limit}, the $\pi N$ scattering amplitude 
can be expressed as a sum of two contributions \cite{wei-nn}.
i) Through the gradient coupling of the pion (referred to as the
$\pi$ vertex), the nucleon could be a potential non-local source ( Fig.2a ) 
for $\sigma' \sim :\vec \pi^2:$.Thus the simplest box and the crossed
box diagrams could effectively produce a $\sigma$ channel interaction.
To extract this contribution we compute the coupling to the composite
$\pi^2$ operator (Fig.2a).As a consequence of the gradient coupling of
the pion, this triangle contribution vanishes 
linearly as $ q \rightarrow 0$.This means the $\sigma'$ decouples from
the nucleon in the low momentum limit.  This is a general result in the 
chiral limit as seen below using the effective Lagrangian approach.
i) There is an effective local vertex, referred to as the $\rho$ vertex,
(Fig.2b) with the pions in spin=1, isospin=1 channel.This is irrelevant
for the $\sigma$ channel.

\begin{figure}[htb]
\begin{center}
\mbox{\epsfig{file=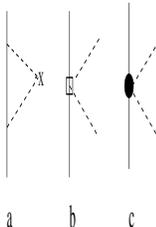,width=3truecm,height=2truecm,angle=-90}}
\caption{ Contributions to the $\pi$ N scattering at low energies.
a.The nucleon could be a non-local source for the $\sigma$ channel.
b.The `$\rho$ vertex' irrelevant for the $\sigma$ channel.
c.The `$\Sigma$ vertex' coming entirely from explicit breaking of 
chiral symmetry.}
\label{Fig 2.}
\end{center}
\end{figure}
In the linear $\sigma$-model the Lagrangian density involving the nucleons 
in the chiral limit, is \cite{wei-nl}, 
\begin{equation}
L_N = \overline N i \gamma^\mu \partial_\mu N - g 
\overline N ( \sigma +i \gamma_5 \vec \tau.\vec \pi)N
\end{equation}
It appears that $\overline N N$ is a source for $\sigma'$.
If so its exchange would give an unusual long ranged nuclear force, 
Eqn.\ref{sigir} in the chiral limit, and would be a model independent 
consequence of the chiral symmetry.By a redefinition of fields 
$L_N$ can be rewritten as \cite{wei-nl}
\begin{eqnarray}
L_N & = & \overline N'(i\gamma^\mu \partial_\mu - 
g (\sigma^2+\vec \pi^2)^{1/2})N' \nonumber \\
& & +  \overline N' \gamma^\mu (1+\frac{\vec \pi^2}{4f^2})^{-1}
(\gamma_5 \frac{\vec\tau.\partial_\mu \vec \pi} {2f}
-\frac{\vec\tau.\vec\pi \times  \partial_\mu \vec \pi}{4f^2})N' 
\label{ln}
\end{eqnarray}
This gives the same S-matrix elements.Therefore only the 
chiral invariant combination $ (\sigma^2+\vec \pi^2)^{1/2}$
couples to the nucleons in the limit of vanishing momenta.
The other couplings are the derivative coupling of (an odd number of)
pions to the axial current (`$\pi$ vertex')
and to even number of pions in the vector-isovector channel (`$\rho $ vertex').
The $SU(2)_L \times SU(2)_R$ singlet 
\begin{equation}
(\sigma^2+\vec \pi^2)^{1/2} - f \sim   \sigma'+\vec \pi^2/(2f) + .... 
\label{inv}
\end{equation}
does not have infrared divergent correlation functions.For
instance, in the non-linear $\sigma$-model this degree of freedom is 
frozen.Also in the linear $\sigma$-model the infrared behaviour of
the $\sigma'$ is the same as that of $-\vec\pi^2$. Indeed the
underlying reason for the identification of $\sigma'$ with
$-\vec\pi^2$ in the infrared is that the chiral invariant
combination occurring in Eqn (4) does not have any infrared divergence
while the individual terms do have logarthmic divergences( in four
dimensions). Further, this chiral invariant field is expected to be
massive.
In QCD we may expect the chiral invariant field to have a mass of O(1Gev)
and its exchange would generate scalar isoscalar forces with a very 
short range.It is not an explanation for the intermediate
range attraction.  Thus {\it in the chiral limit there are no long
range interactions in the $\sigma$ channel of the nucleon-
nucleon interactions, even though the $\sigma'$-propagator 
is infrared divergent}.The local four (and multi-) nucleon terms 
permitted by chiral symmetry \cite{wei} may be regarded as mediated by exchanges
of such very massive chiral singlet fields.{\it They are small 
not because they are forbidden by chiral symmetry but because 
they are suppressed by inverse powers of the heavy masses.}
Our general derivation explains the cancellations due to chiral
symmetry in one loop calculation observed in Ref. \cite{bal}.

We now argue that the {\it small explicit breaking of chiral symmetry 
is crucial for the intermediate range attraction}.
In the linear $\sigma$-model the effects of the current quark mass 
$m_q$  are mimicked by the term $fm_\pi^2 \sigma$ (with $m_\pi^2 \sim O(m_q)$).
It induces a non-zero pion mass $m_\pi^2 \sim O(m_q)$.In addition we include 
\cite{lyn} a small $\it current$  mass  $\Sigma_N \overline NN$ with $\Sigma_N
=O(m_q)$ for the nucleons, as allowed by the transformation properties.
This is $\it not$ double counting  
as we are considering direct effects of the quark mass on two distinct 
sectors of the effective theory, the mesons and the baryons.
It has been pointed \cite{lyn} that canonical PCAC 
relation is not valid in the presence of the $\Sigma_N$ term. 
This is not a problem as we are not using PCAC  for calculating the
amplitudes.  We are explicitly putting in the information that the
divergence of the axial vector current acts not only as an interpolating
field for the pion but also has a non-zero matrix element between
baryons of opposite chirality.

Making the Dyson-Weinberg transformation \cite{wei-nl} for
the combination $((\sigma +g^{-1} \Sigma_N)+i \gamma_5 \vec \tau.\vec \pi) $
we get the Yukawa interaction,
\begin{equation}
\sim
\overline N'N'((\sigma ^2+ \vec \pi^2)+2 g^{-1}\Sigma_N \sigma 
+ g^{-2}\Sigma_N^2)^{1/2} 
\end{equation}
Now there is a direct coupling of $\overline N'N'$ to $\sigma'$ 
in addition to the coupling to the chiral invariant
combination $(\sigma^2 + \vec \pi^2)^{1/2}$.This induces a scalar 
isoscalar force of range much larger than the microsopic scale.
This is also seen in the non-linear $\sigma$ model where the two
chiral symmetry breaking terms are represented by \cite{lyn},
\begin{equation}
L'_N  = -\frac{1}{2} m_\pi^2 \frac{\vec \pi^2}{1+ \vec \pi^2 /(4f^2)}
(1-\frac{\Sigma_{N}\overline N'N'}{f^2 m_\pi^2})
\end{equation}
as allowed by the chiral transformation properties.The second term, 
referred to as the $\Sigma$ vertex (Fig. 2c), is the well known pion nucleon 
$\Sigma$ term, important for the pion nucleon  
scattering length.It shows that there is a  
direct coupling of the nucleon to $\vec \pi^2$ (equivalently to 
$\sigma'$), once the chiral symmetry in explicitly broken. This induces 
the intermediate range attraction.When chiral symmetry is explicitly broken, 
$\sigma'$ propagator is 
no longer infrared singular.The leading behaviour for small $m_q$
is simply obtained by using a massive propagator for $\pi$ in Fig. 1b.
We get the amplitude in the scalar isoscalar channel for a small $m_q$ to be,
\begin{equation}
\frac{3\Sigma_{N}^2}{2f^4}  
\int \frac{d^4 k}{(2\pi)^4} \frac{1}{(k^2-m_\pi^2) ((k+q)^2- m_\pi^2)}
\end{equation}
where a ultraviolet cutoff is presumed.This gives the leading
non-analytic (in $m_\pi$ and $q$) to be 
\begin{equation}
i \frac{3\Sigma_{N}^2}{32\pi^2f^4}  
\int\limits^1_0 dx~~ ln \; \frac{\Lambda^2}{m_\pi^2-x(1-x)q^2}
\label{lead}
\end{equation}

Thus the longest range interaction due to a two
pion exchange is also the most dominant due to the logarithmic
enhancement and is entirely due to explicit breaking of
chiral symmetry.An expansion in powers of $q^2$ involve inverse powers of
$m_\pi^2$ and therefore large.In contrast,  
exchanges of four or more pions are not infrared divergent and not
universal.They are suppressed by inverse powers of the physical cut off,
and therefore numerically small.We emphasise that the Eqn. \ref{lead}
is the dominant contribution for $q << m_\pi$.For larger $q$, there are
corrections from multipion exchanges and because the nucleon is a
non-local source of pions. These corections can be systemmatically
computed using a modification of the chiral perturbation theory \cite{ahs}.

Contribution  of $\Sigma$ vertex has  been considered 
earlier in Ref. \cite{del} in a related context.
We emphasise that we have extracted this as the dominant behaviour
to all orders.
\begin{figure}[htb]
\begin{center}
\mbox{\epsfig{file=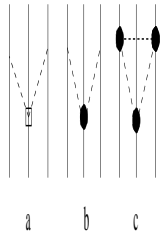,width=3truecm,height=2truecm,angle=-90}}
\caption{ Dominant contributions to the three body forces comes from
a.one ` $\rho$ vertex' and two ` $\pi$' vertices
b.one ` $\Sigma$ vertex' and two ` $\pi$' vertices
c.three ` $\Sigma$ vertices'.} 
\label{Fig 3.}
\end{center}
\end{figure}
We now address the implications for the multinucleon interactions.  
We restrict ourselves to the non-linear $\sigma$ model as the 
linear $\sigma$ model gives identical results as with the NN scattering.
We focus on the three
nucleon forces here.In the chiral limit only Fig.3a involving one $\rho$
-vertex and two $\pi$ vertices survives and gives rise to a long range 
interaction.This contributes only for nnp or 
ppn systems(related to the vanishing coupling $\rho^0 \pi^0 \pi^0$).
In the non-relativistic limit where the energy transfer is zero,
each of the three vertices  is suppressed by a power of the momentum transfer.
With an explicit breaking of chiral symmetry, 
there are contributions from one $\Sigma$ and two $ \pi$ vertices (Fig.3b) 
and from three $\Sigma$ vertices (Fig.3c).We estimate the three contributions
as $q^3/m_\pi^4, \Sigma_N  q^2/m_\pi^4$ and $\Sigma_N^3/m_\pi^2$ 
respectively.In a theory like QCD we expect  
$\Sigma_N \sim m_\pi^2  \sim m_q$.Hence for $ q^2 << m_\pi^2$ Fig
3c is leading order in $q^2$ but quadratic in $\Sigma_N$ while Fig 3b
is of order $\Sigma_N$ but suppressed by one power of $q^2/m_\pi^2$.
We emphasise that these contributions are unavoidable and computable 
consequences of chiral symmetry.All other contributions are suppressed
by the heavy masses. Analogues of Fig 3c are also present for
multinucleon forces and they are all of order $\Sigma_N^2$.

In the linear $\sigma$ model, $\sigma$ appears as a fundamental field,
and an exchange of a dressed $\sigma'$ would appear as a tree diagram in
the skeleton expansion.This is in contrast to the non-linear $\sigma$ 
model where it would appear as a loop effect.As the $\sigma$ channel 
plays a crucial role for nuclear forces it is more appropriate to
use the formalism of the linear $\sigma$ model.On the other hand it 
would appear that the consequences of the low energy theorems are 
reflected better in the non-linear $\sigma$ model.The formalism of
$1/N$ expansion \cite{abhs} \cite{ahs} provides a unified formalism 
combining the best of both.

In this  Letter we have argued that because of the chiral symmetry 
it is possible to analyse exhaustively the dominant effects of
pion exchanges between nuclei in a model independent way.
The $\sigma$ `particle'  that 
effectively mediates the scalar isoscalar channel has a
model independent and unusual logarithmic infrared behaviour.However 
in the chiral limit this decouples from the nucleon at low
momenta.Consequently there are no long range interactions in this 
channel. Thus the intermediate range attraction is much weakened 
because of chiral symmetry and it is the `small'  explicit breaking 
which governs it.  Also the $\sigma'$ is equivalent to $:\vec \pi^2:$ 
in the infrared in a model independent way.As a result the intermediate 
range attraction is dominated by a very simple process: there is an 
effective $\overline NN :\vec \pi^2:$ vertex
whose strength is fixed by the $\pi N$ elastic scattering, and is
theoretically of the order of the quark mass.The two pion exchange   	
between two nucleons induced via this effective vertex summarises 
the dominant effect of all kinds of virtual exchanges between
the nucleons. Thus naive perturbation theory considerations turn
out to be correct for not so naive reasons.Also there are small
and computable multinucleon forces of intermediate range.Here we
have considered only the dominant contributions for $q<<m_\pi$.
It is possible to systemmatically compute \cite{ahs} corrections
of $O(q^2)$ and higher by a modification of chiral perturbation 
theory and thereby extend the range of validity to $q \sim m_\pi$.
Nevertheless very accurate data for NN and Nd scattering in the 
range of few MeV exists, for which the calculations here are directly 
relevant.


\begin{references}
\bibitem[\ddag]{email}{\em electronic address :}
ramesha,dass,sharat@imsc.ernet.in
\bibitem{wei} S. Weinberg,Phys. Lett.{\bf B251},288,(1990);
Nucl. Phys.{\bf B363},3,(1991).
\bibitem{wei-nl} S. Weinberg,Phys. Rev. Lett.{\bf 18},188,(1967).
\bibitem{wei-nn} S. Weinberg,Phys. Rev. Lett.{\bf 17},616,(1966).
\bibitem{cpt} J.Gasser,H.Leutwyler, Phys.Repts. {\bf 87},77,(1982);
U.Meissner, Rep.Prog. Phys.{\bf B56},903(1993); H.Leutwyler,{\it Recent 
Aspects of Quantum Fields},ed. H.Mitter and M.Gausterer, Lecture Notes in
Physics {\bf 396}(1991)1; A.Pich, Rep.Prog. Phys.{\bf B58},563(1995);
G.Ecker,Chiral Perturbation Theory,UWThPh-1996-34,June 1996
hep-ph/9608226; 
\bibitem{cel} C.Celenza, A.Pantziris and C.M.Shakin, Phys. Rev.{\bf C46},2213,(1992);
\bibitem{bir} Birse, Phys. Rev.{\bf C49 },2212,(1994);
\bibitem{de} deRocha and M.R.Robilotta, Phys. Rev.{\bf C49 },1818,(1994);
\bibitem{fri} J.L.Friar and S.A.Coon, Phys. Rev.{\bf C49 },1272,(1994);
\bibitem{park} Park T-S, Min and Rho M.,Phys. Repts..{\bf B233},341,(1993);
\bibitem{ordo} C.Ordonez and V.Van Kolck, Phys.Lett.{\bf B291},459,(1992);
C.Ordonez L.Ray and V.Van Kolck, Phys. Rev.Lett.{\bf 72},1982 (1994);
Phys.Rev.C53,2086(1996).
\bibitem{bal} J-L. Ballot, M.L.Robilotta and C.A.da Rocha, nucl-th/9611.
\bibitem{del} J.Delorme, G.Chanfray and M.Erickson, Nucl. Phys.
{\bf A603},239(1996). 
\bibitem{kap} D.B.Kaplan,M.J.Savage and M.B.Wise,nucl-th/9605002(1996).
\bibitem{rg} 
E. Brezin, D.J. Wallace and K.G. Wilson, Phys. Rev.  {\bf B7}, 232(1973);
D.J. Wallace and R.K.P. Zia, Phys. Rev. {\bf B12},5340(1975);
E. Brezin and J.Zinn-Justin, Phys. Rev.  Lett.{\bf 36} (1976) 691.
\bibitem{abhs} Ramesh Anishetty, Rahul Basu, N.D.Hari Dass and 
H.S.  Sharatchandra, imsc-94/52, hep-th/9502003.
\bibitem{ahs} Ramesh Anishetty, N.D.Hari Dass and H.S.  Sharatchandra, 
under preparation.
\bibitem{leh} H.Lehmann , Phys. Lett. {\bf B41},529(1972).
\bibitem{lyn} B.W.Lynn, Nucl. Phys.{\bf B402},281(1993).
\end{references}
\end{document}